\documentclass{elsart}

\def\endproof{\vrule height6pt width6pt depth0pt}

\begin{document}

\begin{frontmatter}

\title{Recursive proof of the Bell-Kochen-Specker theorem in any dimension $n>3$}
\author{Ad\'{a}n Cabello}
\ead{adan@us.es}
\address{Departamento de F\'{\i}sica Aplicada II,
Universidad de Sevilla, E-41012 Sevilla, Spain}
\author{Jos\'{e} M. Estebaranz}
\address{Departamento de F\'\i sica Te\'orica I,
Universidad Complutense, E-28040 Madrid, Spain}
\author{Guillermo Garc\'{\i}a-Alcaine}
\ead{ggarciaa@fis.ucm.es}
\address{Departamento de F\'\i sica Te\'orica I,
Universidad Complutense, E-28040 Madrid, Spain}


\begin{abstract}
We present a method to obtain sets of vectors proving the
Bell-Kochen-Specker theorem in dimension~$n$ from a similar set in
dimension $d$ ($3\leq d<n\leq 2d$). As an application of the
method we find the smallest proofs known in dimension five
(29~vectors), six (31) and seven (34), and different sets matching
the current record (36) in dimension eight.
\end{abstract}



\end{frontmatter}


\section{Introduction}
\label{sec:I}


The Bell-Kochen-Specker (BKS) theorem~\cite{Bell66,KS67} states
that quantum mechanics (QM) cannot be simulated by noncontextual
hidden-variable theories. Any hidden-variable theory reproducing
the predictions of QM must be {\em contextual} in the sense that
the result of an experiment must depend on which other compatible
experiments are performed jointly. The BKS theorem is independent
of the state of the system, and is valid for systems described in
QM by Hilbert spaces of dimension $d \ge 3$.

A proof of the BKS theorem consists of a set of physical yes-no
tests, represented in QM by one-dimensional projectors, to which
the rules of QM do not allow the assignment of predefined ``yes''
or ``no'' answers, regardless of how the system was prepared. In
this Letter, yes-no tests will be represented by the vectors onto
which the projectors project.

Several proofs of the BKS theorem in dimensions three, four and
eight are known: see, for instance,~\cite{CEG96} and the
references in~\cite{CG96}. General procedures for extending the
demonstration to a finite dimension~$n$ also
exist~\cite{CG96,ZP93,Peres95}. In Section~\ref{sec:II}, we
present a new method to obtain sets of vectors proving the BKS
theorem in dimension~$n$ from a similar set in dimension~$d$
($3\leq d<n\leq 2d$). In Section~\ref{sec:III} we compare this
method with those of~\cite{CG96,ZP93,Peres95}. The main interest
of this method is that it leads to the smallest proofs known in
dimension five (29~vectors), six (31) and seven (34), and to
different sets matching the current record (36) in dimension
eight. These proofs are explicitly presented for the first time in
Section~\ref{sec:IV}; a preliminary version of them was referred
to in~\cite{Bub97,Cabello00}.

Which one is the smallest number of yes-no tests needed to prove
the BKS theorem in each dimension? This is an old
question~\cite{CG96}. Recently, it has been proven that the answer
is 18 for dimension four~\cite{PMMM05}, and that there are no
proofs with less yes-no tests in any dimension~\cite{CPP05}. The
proofs presented in Section~\ref{sec:IV} give an upper bound to
this search in dimensions five to eight. The important point is
that these bounds are sufficiently small so as to apply recently
developed approaches capable to exhaustively explore all possible
proofs of the BKS theorem~\cite{PMMM05,CPP05}. The practical
limitation of these approaches is that the complexity of the
exploration grows exponentially with the number of vectors, making
it difficult to explore all possible sets involving 30 vectors or
more.

A set of $n$-dimensional vectors $X:=\left\{{\bf
u}_j\right\}_{j=1}^N$ is a proof of the BKS theorem if we cannot
assign to each vector ${\bf u}_j$ a $v({\bf u}_j)$ such that:

\begin{itemize}
\item[(a)] Each $v({\bf u}_j)$ has a uniquely defined value, 0
or~1 (``black'' or ``white''); this value is {\em non-contextual},
i.e., does not depend on which others $v({\bf u}_k)$ are jointly
considered.

\item[(b)] $\sum\limits_{i=1}^nv({\bf u}_i) =1$ $\forall$ set
of~$n$ mutually orthogonal vectors $\left\{{\bf
u}_i\right\}_{i=1}^n\in X$.
\end{itemize}

In that case $X$ is said to be ``non-colourable''. A proof of the
BKS theorem is said to be ``critical'' if all vectors involved are
essential for the proof.


\section{Recursive proof of the Bell-Kochen-Specker theorem}
\label{sec:II}


Let $A:=\left\{ {\bf a}_i\right\} _{i=1}^f$, ${\bf
a}_i=(a_{i1},\ldots ,a_{id})$, be a proof in dimension $d$. For
any $n:=d+m$, $1\leq m\leq d$, let us define two sets of
$n$-dimensional vectors, $B^{*}:=\left\{ {\bf b}_i\right\}
_{i=1}^f$, $C^{*}:=\left\{ {\bf c}_i\right\} _{i=1}^f$, obtained
by appending to each vector ${\bf a}_i$ $m$ zero components {\em
on the right} and {\em on the left}, respectively; ${\bf
b}_i:=(a_{i1},\ldots ,a_{id},0,\ldots ,0) $, ${\bf
c}_i:=(0,\ldots,0,a_{i1},\ldots ,a_{id}) $. Let us also define the
following sets of $n$-dimensional vectors: $\overline{B}:=\left\{
{\bf b}_j\right\} _{j=f+1}^{f+m}$, $b_{jk}:=\delta _{j-f+d,k}$;
$\overline{C}:=\left\{ {\bf c}_j\right\} _{j=f+1}^{f+m}$,
$c_{jk}:=\delta _{j-f,k}$; $B:=B^{*}\cup \overline{B}=\left\{ {\bf
b}_j\right\} _{j=1}^{f+m}$, $C:=C^{*}\cup \overline{C}=\left\{
{\bf c}_j\right\} _{j=1}^{f+m}$.

{\bf Lemma.} {\em $B$ is BKS-colourable if and only if}
\begin{equation}
\label{e1}\sum\limits_{j=f+1}^{f+m}v({\bf b}_j) =1.
\end{equation}

{\bf Proof.} The sets of $d$ mutually orthogonal vectors in $A$
become sets of $n$ mutually orthogonal vectors in $B$, sharing the
last $m$ vectors, ${\bf b}_j\in \overline{B}$, $j=f+1,\ldots,
f+m$. If condition (\ref{e1}) is fulfilled, we can colour $B$
simply by assigning the values $v({\bf b}_j)=0$, $j=1,\ldots, f$;
conditions (a) and (b) are automatically satisfied. Conversely:
if~(\ref{e1}) is not verified, then $v({\bf b}_j) =0$,
$j=f+1,\ldots, f+m$; the impossibility to colour set $A$ in
dimension $d$ following rules (a), (b) implies the impossibility
to colour $B$ in dimension $n$. \hfill\endproof

The same reasoning applies to $C$: $C$ is colourable if and only if
\begin{equation}
\label{e2}\sum\limits_{j=1}^mv({\bf c}_j) =1.
\end{equation}

{\bf Theorem.} {\em $D:=B\cup C$ is a non-colourable set.}

{\bf Proof.} If $d<n\leq 2d$, then $\overline{B}\cap
\overline{C}=\emptyset$; conditions (\ref{e1}) and (\ref{e2}),
necessary to colour $B$ and $C$, would imply the existence of two
mutually orthogonal vectors, ${\bf b}_k\in \overline{B}$, ${\bf
c}_l\in \overline{C}$, with values $v({\bf b}_k) =1$, $v({\bf
c}_l) =1$; this prevents $D=B\cup C$ from being coloured following
rule (b); therefore $D$ is a non-colourable set. \hfill\endproof

The number $g$ of different vectors in set $D$ is $g\leq 2(f+m)$;
the extreme is reached only if $B\cap C=\emptyset $. In general,
set $D$ is not critical (i.e., some subsets of $D$ are
also non-colourable sets). To search for critical subsets, we will
use a generalization to arbitrary dimension of the computer
program of Ref.~\cite{Peres95p209}.


\begin{table}
\begin{center}
\begin{tabular}{p{2.33 in} |cccc}
\hline
Dimension $n$ & 5 & 6 & 7 & 8 \\
\hline
 & & & & \\
$n\leq 2d$ method, starting from Peres' 24-vector set in
$d=4$~\cite{Peres91} & 39 ({\bf 29}) & 44 ({\bf 31}) & 47 ({\bf 34}) & 48 ({\bf 36}) \\
 & & & & \\
$n\leq 2d$ method, starting from the 18-vector critical set $S_4$
in $d=4$ & 31 ({\bf 29}) & 35 (32) & 37 ({\bf 34}) & 38 ({\bf 36}) \\
 & & & & \\
Zimba-Penrose method~\cite{ZP93}, using Conway and Kochen's 31-vector
critical set in $d=3$~\cite{Peres95p114} and $S_4$ & \ldots & 62 & 49 & {\bf 36} \\
\hline
\end{tabular}
\end{center}
\vspace{0.4cm} \noindent TABLE 1. {\small Number of vectors of
some proofs of the BKS theorem in dimensions five to eight
obtained by several methods. In parenthesis are the sizes of the
smallest critical subsets obtained by computer (see
Section~\ref{sec:IV} for examples of such sets). Records in each
dimension are in boldface.}
\end{table}


\section{Comparison with other methods}
\label{sec:III}


In the following, we will present some outcomes of the ``$n\leq
2d$ method'' introduced in Section~\ref{sec:III}, and compare them
with those obtained with other methods.

The $n\leq 2d$ method allows us to construct non-colourable sets
in any dimension $n\geq 4$, starting with one non-colourable set
in dimension three (first in dimension four, five, and six; them
up to dimension~12 using the sets generated in this first step,
etc.): we could start with Conway and Kochen's 31-vector critical
set, reviewed in~\cite{Peres95p114}, or with Peres' 33-vector
critical set~\cite{Peres91}; nevertheless, smaller sets in
dimension $n\geq 5$ can be obtained starting from suitable
non-colourable sets in dimension $n=4$~\cite{CEG96,Peres91}. In
general, starting with a $f$-vector non-colourable set in
dimension~$d$, this method produces non-colourable sets with {\em
at most} $g\leq 2^k(f+kd) $ vectors in dimension~$2^kd$.

The first row in Table~1 shows the number of vectors of the proofs
in dimensions five to eight obtained starting from the
non-critical 24-vector set of Peres' (P-24 in the
following)~\cite{Peres91} in dimension~$n=4$. In parentheses are
the sizes of their smallest critical subsets found by means of a
computer search. For dimensions $n=5,6,7$ these are the current
records (denoted in boldface). In dimension $n=8$, the 48-vector
set obtained by this method contains 256~different 36-vector
critical subsets matching the record of Kernaghan and
Peres'~\cite{KP95}. Record non-colourable sets in dimensions four
to eight are presented in Section~\ref{sec:IV}.

We could also start from the 18-vector critical set $S_4$ given in
Section~\ref{sec:IV}: the number of vectors of the resulting
non-colourable sets, shown in the second row of Table~1, are
reasonably close to the record in each dimension; the sizes of the
smallest critical subset are shown in parentheses.

Composing two non-colourable sets with $f$ and $g$ vectors in
dimensions~$d$ and $m$, Zimba and Penrose's (ZP's)
method~\cite{ZP93} produces a non-colourable set with $f+g$ in
dimension~$d+m$, which is critical if both components are
critical. ZP's method leaves out the case $n=5$, since both $d$
and $m$ must be greater or equal than three. The number of vectors
increases linearly if the dimension is a multiple of the initial
one; starting with a $f$-vector critical set in dimension~$d$,
ZP's method produces {\em critical} sets with $hf$ vectors in
dimension $hd$ (note that, although {\em these} sets do not
contain any subset that is also a non-colourable set, the number
$hf$ is only an upper bound to the size of the smallest possible
non-colourable sets in dimension $hd$). In particular, starting
from the 18-vector sets of Ref.~\cite{CEG96}, ZP's method will
produce $2^n$ different critical sets with $9\times 2^{k+1}$
vectors in dimension $n=2^{k+2}$, $k=1,2,\ldots$, compared with a
non-critical $g$-vector non-colourable sets with $9\times
2^{k+1}<g\leq (9+2k) \times 2^{k+1}$ obtained by our previous
$n\leq 2d$ method.

The third row in Table~1 represents the number of vectors of the
non-colourable sets obtained with the ZP method, starting with the
smallest non-colourable sets currently known in three and four
dimensions (Conway and Kochen's 31-vector critical
set~\cite{Peres95p209}, and any of the 18-vector critical sets of
Ref.~\cite{CEG96}, respectively). The non-colourable sets in
dimensions six and seven are larger than those previously
discussed (and do not contain smaller non-colourable subsets,
because they are critical). In dimension $n=8$, starting from
couples of the 18-vector sets of Ref.~\cite{CEG96} ZP's method
produces 256~different critical 36-vector sets (which, actually,
are the same 256~record sets obtained with the~$n\leq 2d$ method).

Other methods produce larger non-colourable sets: for instance,
in~\cite{Peres95p212} Peres constructs a proof in dimension $d+1$
starting from another in dimension~$d$; only one initial
non-colourable set is needed to reach recursively any
dimension~$n$, but the sizes of the non-colourable sets obtained
increase rapidly in general, and a search for critical subsets is
necessary in order to avoid very large sets. Finally, the method
in Ref.~\cite{CG96} is a generalization to arbitrary dimension of
the three-step original construction of Kochen and
Specker's~\cite{KS67}, explicitly showing the relation between the
different kinds of non-colourable sets and the three versions of
the BKS theorem, but it is almost as inefficient as regards the
number of vectors involved (96 in dimension $n=3$, 136 in $n=4$,
and $35n$ if $n\geq 5$) as the original Kochen-Specker 117-vector
proof in dimension $n=3$ was.


\section{Record critical proofs in dimensions four to eight}
\label{sec:IV}


As an application of the method introduced in Section~\ref{sec:II}
we have obtained the smallest proofs known in dimension five
(29~vectors), six (31) and seven (34), and several sets matching
the current record (36) in dimension eight. We have omitted the
normalization constants of the vectors in order to simplify the
notation.

Let us start with an example~\cite{CEG96} of the smallest proof of
the BKS theorem, not only in dimension four~\cite{PMMM05}, but in
any dimension~\cite{CPP05}: $S_4:=\{(1,0,0,0)$, $(0,0,1,0)$,
$(0,0,0,1)$, $(1,1,0,0)$, $(0,1,1,0)$, $(0,0,1,1)$, $(1,-1,0,0)$,
$(0,1,-1,0)$, $(1,0,1,0)$, $(0,1,0,1)$, $(0,1,0,-1)$, $(1,0,0,1)$,
$(1,-1,1,-1)$, $(1,1,-1,-1)$, \linebreak $(1,-1,-1,1)$,
$(1,1,1,-1)$, $(1,1,-1,1)$, $(-1,1,1,1)\}$. $S_4$ is one of the
16~similar sets of Ref.~\cite{CEG96}. We can construct 9~tetrads
of mutually orthogonal vectors in terms of the 18~elements of
$S_4$; each vector is orthogonal to 7~others in the set and
appears in 2~tetrads. The non-colourability of the set is proved
by a parity argument: the sum of values for each tetrad is 1,
following (b); therefore the sum of the values in the 9~tetrads
must be 9, but each value appears twice and therefore the sum is
{\em even,} following (a). We have chosen a different set than
in~\cite{CEG96} because the proof in $n=5$ deduced from that set
has 33~vectors, instead of 31 as when starting from $S_4$.

An example of the smallest known proof in dimension five is
$S_5:=\{({\bf a},0)$, $(0,{\bf a}):{\bf a}\in
S_4\}-\{(0,1,0,0,0)$, $(0,0,1,0,0)\}$. We can construct
16~pentads of mutually orthogonal vectors in terms of the
29~elements of $S_5$. The non-colourability of $S_5$ can be proved
by exhaustive computer tests (we have found no analytic proof of
the non-colourability of this or the following $S_6$, $S_7$ sets).
This is one of the 120~similar 29-vector critical subsets in the
39-vector non-colourable set obtained from P-24.

An example of the smallest known proof in dimension six is
$S_6:=\{({\bf a},0,0)$, $(0,0,{\bf a}):{\bf a} \in S_4\}$ $\cup$
$\{(0,1,0,0,0,0)$, ${\bf (1,0,-1,0,0,0)}$, ${\bf (1,1,1,1,0,0)}\}$
$-$ $\{(0,0,1,0,0,0)$, $(0,0,0,1,0,0)$, $(1,1,0,0,0,0)$,
$(0,0,1,-1,0,0)$, \linebreak $(1,-1,-1,1,0,0)$, $(0,1,0,1,0,0)\}$;
vectors in boldface appear when applying the $n\leq 2d$ method to
P-24, but not to $S_4$. We can construct 16~hexads of mutually
orthogonal vectors in terms of the 31~elements of $S_6$. The
non-colourability of $S_6$ has been proved by computer. This is
one of the 128~similar 31-vector critical subsets in the 44-vector
non-colourable set obtained from P-24.

An example of the smallest known proof in dimension seven is
$S_7:=\{({\bf a},0,0,0)$, $(0,0,0,{\bf a}):{\bf a}\in
S_4\}$$-\{(0,0,0,1,0,0,0)\}$. We can construct
28~heptads of mutually orthogonal vectors in terms of the
34~elements of $S_7$. The non-colourability of $S_7$ has been
proved by computer. This is one of the 144 similar 34-vector
critical subsets in the 47-vector non-colourable set obtained from
P-24.

An example of the smallest known proof in dimension
eight~\cite{KP95} is \linebreak $S_8:=\{({\bf a},0,0,0,0)$,
$(0,0,0,0,{\bf a}): {\bf a}\in S_4\}$. We can construct 81~octads
of mutually orthogonal vectors in terms of the 36~elements of
$S_8$; each vector is orthogonal to 25~others in the set and
belongs to 18~octads. The non-colourability of $S_8$ can be proved
by a parity argument: the number of octads is {\em odd}, but each
vector appears an {\em even} number of times (actually, the
non-colourability of $S_8$ is a consequence of the
non-colourability of $S_4$ and the application of the ZP method).
$S_8$ is one of the 256~similar 36-vector critical subsets in the
48-vector non-colourable set obtained from P-24.


\section*{Acknowledgments}


We would like to thank the late Asher Peres for his comments and
advice.



\begin{thebibliography}{00}

\bibitem{Bell66}
J.S.~Bell,
Rev. Mod. Phys. 38 (1966) 447.

\bibitem{KS67}
S. Kochen and E.P.~Specker,
J. Math. Mech. 17 (1967) 59.

\bibitem{CEG96}
A. Cabello, J.M.~Estebaranz, G.~Garc\'\i a-Alcaine,
Phys. Lett.~A 212 (1996) 183.

\bibitem{CG96}
A. Cabello, G.~Garc\'\i a-Alcaine,
J. Phys. A 29 (1996) 1025.

\bibitem{ZP93}
J. Zimba, R. Penrose,
Stud. Hist. Phil. Sci. 24 (1993) 697.

\bibitem{Peres95}
A. Peres,
Quantum Theory: Concepts and Methods,
Kluwer, Dordrecht, 1995.

\bibitem{Bub97}
J. Bub,
Interpreting the Quantum World, Cambridge University Press,
Cambridge, 1997, p.~119.

\bibitem{Cabello00}
A.~Cabello,
Int. J. Mod. Phys.~A 15 (2000) 2813.

\bibitem{PMMM05}
M.~Pavi\v{c}i\'{c}, J.-P.~Merlet, B.D.~McKay, N.D.~Megill,
J. Phys.~A 38 (2005) 1577.

\bibitem{CPP05}
A.~Cabello, J.R.~Portillo, G.~Potel,
unpublished.

\bibitem{Peres95p209}
Ref.~\cite{Peres95}, p.~209.

\bibitem{Peres95p114}
Ref.~\cite{Peres95}, p.~114.

\bibitem{Peres91}
A. Peres,
J. Phys.~A 24 (1991) L175.

\bibitem{KP95}
M. Kernaghan, A. Peres,
Phys. Lett.~A 198 (1995) 1.

\bibitem{Peres95p212}
Ref.~\cite{Peres95}, p.~212.

\end{thebibliography}
\end{document}